\documentclass[journal]{IEEEtran}

\usepackage{booktabs, multicol, multirow}
\usepackage{graphicx, amsfonts,amsmath , boldline, epstopdf,amssymb }
\usepackage{tabu, multirow,multicol ,booktabs ,setspace ,algcompatible ,mathrsfs,dsfont, array, boldline, makecell, booktabs }
\usepackage{url}

\usepackage{epstopdf}
\usepackage{adjustbox, stackengine, color, colortbl, epsfig,cite, cases, enumitem, mathtools, tikz, verbatim}
\newcolumntype{?}{!{\vrule width 1.2pt}} % for thicker the vertical line
\usepackage[caption = false]{subfig}
\newcommand{\tabincell}[2]{\begin{tabular}{@{}#1@{}}#2\end{tabular}}
\usepackage[utf8]{inputenc}
\newcommand\witwo{1.8} % for define the widthof figure
\newcommand\hetwo{1.32} % for define the widthof figure

\begin{document}

\definecolor{um}{rgb}{0.0824, 0.1294, 0.4196}
\definecolor{Abi}{rgb}{0.059, 0.322, 0.729}
\definecolor{orange}{rgb}{1, 0.347, 0}
\definecolor{c}{rgb}{0, 1, 1}
\definecolor{m}{rgb}{1, 0, 1}
\title{\vspace{-1 cm}Investigating the Spatiotemporal Charging Demand and Travel Behavior of Electric Vehicles Using GPS Data: A Machine Learning Approach}

%\normalsize{

% \author{\IEEEauthorblockN{Sina Baghali}
% \IEEEauthorblockA{Civil, Environmental, and Construction\\
% Engineering\\
% University of Central Florida\\
% Orlando, Florida, USA\\
% baghalisina@knights.ucf.edu}
% \and
% \IEEEauthorblockN{Zhaomiao Guo}
% \IEEEauthorblockA{Civil, Environmental, and Construction\\ Engineering\\
% University of Central Florida\\
% Orlando, Florida, USA\\
% guo@ucf.edu}
% \and
% \IEEEauthorblockN{Samiul Hasan}
% \IEEEauthorblockA{Civil, Environmental, and Construction\\ Engineering\\
% University of Central Florida\\
% Orlando, Florida, USA\\
% samiul.hasan@ucf.edu}
% }

\author{\IEEEauthorblockN{Sina Baghali, \textit{Graduate Student Member, IEEE}, Zhaomiao Guo, Samiul Hasan}\\
\IEEEauthorblockA{Civil, Environmental, and Construction Engineering, University of Central Florida, Orlando, FL, USA\\
baghalisina@knights.ucf.edu, guo@ucf.edu, samiul.hasan@ucf.edu}}

\maketitle
\thispagestyle{empty}
\pagestyle{empty}
% \vspace{-2cm}
\begin{abstract}
The increasing market penetration of electric vehicles (EVs) may change the travel behavior of drivers and pose a significant electricity demand on the power system. Since the electricity demand depends on the travel behavior of EVs, which are inherently uncertain, the forecasting of daily charging demand (CD) will be a challenging task. In this paper, we use the recorded GPS data of EVs and conventional gasoline-powered vehicles from the same city to investigate the potential shift in the travel behavior of drivers from conventional vehicles to EVs and forecast the spatiotemporal patterns of daily CD. Our analysis reveals that the travel behavior of EVs and conventional vehicles are similar. Also, the forecasting results indicate that the developed models can generate accurate spatiotemporal patterns of the daily CD.
 
\end{abstract}


%  In addition, countries like the Netherlands and Germany, as well as several U.S. states have announced aggressive plans to reach 100\% zero-emission vehicles in the coming years.

\allowdisplaybreaks
\vspace{-0.5cm}
\section{Introduction}
% \subsection{Background}
The transportation sector is the second-largest source of greenhouse gas emissions (GHGs)\cite{barbosa2018human}. Electric vehicles (EVs), receiving increased popularity, provide an effective way to control GHGs and fossil fuel consumption. With EVs passing 10 million counts in 2020 worldwide \cite{outlook2021electric}, it is foreseeable that EVs will comprise a significant market share of the future transportation fleet, which may cause shifts in travel behavior in the transportation systems and influence the power system operation. According to the International Energy Agency (IEA), the EV fleet will impose 860 TWh of electricity demand by 2030 \cite{outlook2021electric}. Therefore, the spatial and temporal behavior of EV charging demand (CD) is vital for power system planning and operation.  

% In this paper, we will investigate the potential differences in the travel behavior of EVs compared to conventional vehicles. We will also estimate the spatiotemporal pattern of charging demand (CD) based on GPS data.

% EV charging demand is closely related to EV travel behavior. XXX  In this paper, we will investigate the potential differences in travel behavior of EVs compared to the conventional vehicles and estimate the spatio-temporal pattern of charging demand based on emerging data.

% Here we start by providing a review of the data sources used in the studies, and later we will discuss the details of the applied methodologies.
% Survey and consensus data are the classic sources of data used for developing human mobility models. In recent years, however, mobile phone activity data revolutionized human mobility data by providing both geographic and time-stamped data of individual activities. Later on,  Global Positioning System (GPS) data provided even higher accuracy of movement of individuals in space and time. The most recent development in human mobility data is the location information on social networks using the built-in GPS and WiFi chips of smartphones \cite{barbosa2018human}. 
% In the following, we will discuss the mobility data sources used in modeling the travel behavior of EVs.

% \subsection{Literature Review}
The literature in modeling EV CD can be discussed from two aspects: 1) the input data used; and 2) the applied modeling methodologies. Survey data are the classic sources of data used in multiple studies for modeling the driving and charging behavior of EVs \cite{jahangir2019charging,li2018gis,baghali2021analyzing}. However, survey data may be costly to collect and can only provide a perceived travel behavior. Mobile phone and social network data with location tags may not be an appropriate source for analyzing the behavior of EVs because the mode of transportation can not be accurately estimated. Among the mobility data sources, GPS data is the best fit to model and estimate the actual behavior of EVs. With high-resolution location tracking, we can record the driving distance of each trip, calculate the resulting CD of the vehicles, and estimate locations where charging stations are mostly needed. Additionally, the high sample frequency of GPS data can help model the temporal behaviors of EVs accurately. In earlier studies \cite{tian2014understanding,de2014gis}, the GPS data of conventional internal combustion engine vehicles (ICEVs) were considered as EVs and derived charging behaviors based on their travel behaviors. Current studies also use the GPS data of ICEVs to infer the feasibility and potentials of electrifying the existing transportation fleet \cite{yang2017ev,fraile2018using}. These data sources, however, can not account for the behavior of EVs since they are either small-scale data of limited ICEVs or GPS data of specific vehicle types, e.g., taxi fleet \cite{tian2014understanding,yang2017ev,fraile2018using}. 

% A more detailed review on the data sources used in CD modeling is done in \cite{calearo2021review}. 

% Even if they use large-scale data of different ICE types (not just taxis) \cite{de2014gis} they can not represent the characteristics of EVs simply because they are not actual EVs and the resulting charging demand would base on multiple assumptions about their battery capacity and energy consumption rate.

% Several studies consider the data of charging stations to model and forecast the CD of EVs \cite{almaghrebi2020analysis,almaghrebi2020data}. These studies, however, rule out the home-based charging sessions and do not consider the travel behavior of EVs.

% Deterministic modeling is commonly used in determining the CD of EVs, where CD is calculated directly from each trip data and there is no forecasting involved. The results of deterministic modeling, however, can be used to validate the results of the proposed forecasting methods \cite{li2018gis}.

After selecting one of the data sources discussed, studies implement different methodologies to model and estimate the charging and travel behavior of EVs. Stochastic modeling is the most popular method applied in many studies for investigating the stochastic behavior of CD \cite{leou2013stochastic, tang2015probabilistic, arias2016electric,  sun2017novel, li2018gis}. Stochastic modeling considered in the studies are either based on Monte-Carlo Simulation (MCS) \cite{arias2016electric} or Markov chain modeling  \cite{sun2017novel}. Both cases of stochastic modeling require many scenario generations, which is computationally expensive, and they neglect the correlation of travel parameters.

%and machine learning models are the recent alternative tools \cite{panahi2015forecasting,jahangir2019charging,mansour2020machine}

Machine learning algorithms, such as k-nearest neighbors (KNN) \cite{li2018gis} or artificial neural networks \cite{panahi2015forecasting,jahangir2019charging,mansour2020machine}, are recently adopted in CD estimation. However, studies in \cite{panahi2015forecasting,mansour2020machine} incorporate probabilistic models to generate synthetic trips to overcome the small scale of their input data. In \cite{jahangir2019charging}, a large survey data is used for estimating the travel parameters and CD. However, the authors assume smart charging of EV users where a centralized entity determines the optimal time for charging and shifts the charging time to the hours with low charging price, which may not represent the realistic daily CD.  

% Additionally, they have assumed that vehicles will charge only after their last trip of the day, ruling out the possibility of charging during work hours and other public charging instances.
%In summary, the current studies have considered data sources that may introduce unnecessary biases and limitations. Moreover, the proposed modeling methodologies have low estimation accuracy and computation challenges.

% \subsection{Contribution}
In this study, we seek to resolve the aforementioned drawbacks by using GPS data for EVs and developing machine learning models to estimate the daily CD. Unlike 
\cite{yang2017ev,fraile2018using} that considered GPS data of ICEVs as EVs, we have examined the data of actual EVs and provided extensive comparison on the traveling behavior of EVs and ICEVs commuting in the same urban network. Additionally, we have analyzed the CD of EVs based on their recorded state of charge (SOC) without making assumptions on EVs' initial SOC and energy consumption based on the traveling distances, which is a prevalent assumption made in different studies due to the lack of information in the data sets \cite{jahangir2019charging,baghali2021analyzing,mansour2020machine}. Lastly, different from studies like \cite{panahi2015forecasting,jahangir2019charging,mansour2020machine} that have focused on temporal behavior of the CD, we have developed forecasting models to extract and estimate both the spatial and temporal behavior of CD.
 %Previously, studies calculate an estimate of SOCs based on assumptions and driving distance \cite{panahi2015forecasting,jahangir2019charging,li2018gis,baghali2021analyzing}.
% Using the VED data set will allow us to incorporate real energy consumption and SOC of EVs in our model. We will use this dataset for two purposes: 1) Comparing the daily travel pattern of EVs and ICEVs. 2) Extract and estimate the spatiotemporal behavior of CD of EVs.

% \vspace{-0.3cm}
\section{Data Description}
We will use vehicle energy dataset (VED) \footnote{Data: \url{https://github.com/gsoh/VED}}, which is a publicly available data set containing GPS trajectories of a limited number of personal cars including both ICEVs and EVs in Ann Arbor, Michigan, the USA from Nov 2017 to Nov 2018 \cite{oh2020vehicle}. This is a unique data set providing high-resolution data of EVs' energy consumption and their state of charge (SOC). The VED contains trajectories of 383 vehicles, including 264 ICEVs, 92 hybrid EVs (HEVS), and 27 plug-in HEV (PHEVs/EVs). The VED consists of dynamic and static data sets; Static data contains vehicle parameters (e.g., vehicle ID, vehicle type, vehicle class, etc.), and dynamic data contains high-resolution daily trip trajectories and other trip parameters (e.g., day number, trip number, latitude, location latitude, and longitude, etc.). We will derive the vehicle IDs of EVs from static data and use them to extract the dynamic data of EVs based on vehicle ID. %Weekly dynamic data are stored in separate files. We went through the dynamic data files and extracted the trajectories of EVs based on the IDs found from the static data. 
The collected dynamic data has different features of the trips. Among them, we will use day number, vehicle ID, trip number, timestamps (ms),  latitude, longitude, and state of charge (SOC) of batteries. {Table \ref{T:input_data}} shows a sample of the input data with the selected features for an EV as an example.
\begin{table}
\centering
\renewcommand{\arraystretch}{1}
\footnotesize
\captionsetup{labelsep=space,font={footnotesize,sc}}
\caption{Input data example}\label{T:input_data}
\centering
\resizebox{\linewidth}{!}{
\begin{tabular}{|c|c|c|c|c|c|c|}
\hline\hline
{\tabincell{c}{Day \\ No.}} & {\tabincell{c}{Vehicle \\ ID}} & {\tabincell{c}{Trip\\No.}} &
{\tabincell{c}{Time stamp\\(ms)}} & 
{\tabincell{c}{Latitude \\ (deg)}} & {\tabincell{c}{Longitude \\ (deg)}} &
{\tabincell{c}{SOC \\ ($\%$)}} 
\\ [2ex] \hline
\multicolumn{1}{|c|}{5.5602}   & 371 & 1288    & \multicolumn{1}{c|}{0}    & \multicolumn{1}{c|}{42.2776}     & \multicolumn{1}{c|}{-83.7537} & 94.344        \\

\multicolumn{1}{|c|}{5.5602}   & 371   & 1288  & \multicolumn{1}{c|}{600}   & \multicolumn{1}{c|}{42.2776}     & \multicolumn{1}{c|}{-83.7537}  & 94.344       \\

\multicolumn{1}{|c|}{5.5602}   & 371   & 1288  & \multicolumn{1}{c|}{700}   & \multicolumn{1}{c|}{42.2776}     & \multicolumn{1}{c|}{-83.7537}  & 94.344      \\
\multicolumn{1}{|c|}{5.5602}   & 371   & 1288  & \multicolumn{1}{c|}{1700}   & \multicolumn{1}{c|}{42.2776}     & \multicolumn{1}{c|}{-83.7537} & 94.344      \\
\hline\hline
                            
\end{tabular}
}
\end{table}

% \begin{figure}
%   \centering
%   \includegraphics[width=0.5\linewidth]{Figures/trajectory_example.png}
%     \captionsetup{justification=raggedright,singlelinecheck=false}
%   \caption{Yearly trip trajectory example of the input data (vehicle ID = 371) \label{fig:trip_traj_ex}}
% \end{figure}
% In the example provided in {Table \ref{T:input_data}}, EV with ID = 371, starts trip No. 1622 at 22:40:39 on day 45.
% In {Table \ref{T:input_data}}, the day No. is the time of the trips. The whole number represents the day, which can go up to 365, and the fractional part represents the recorded time as a fraction of 24 hours.

We can calculate the trip duration and end time by considering the last timestamp recorded for each trip. Trip distance is another important parameter that can be extracted with the recorded latitudes and longitudes at each timestamp by calculating the distance between each consecutively recorded timestamp and aggregating over the records of each trip using the Haversian formula \cite{khazeiynasab2020resilience}. 
% However, the resolution of the data is very high (at 10 (s) resolution), and the vehicle's location doesn't change much in most cases between two consecutive timestamps. %Therefore, we can calculate the traveled distance more efficiently every $n$ timestamps (e.g., $n=$5) and sum them until the end of the trip. Note that higher values of $n$ will decrease the trip distance accuracy. 
The other key feature of the input data is the recorded SOC of batteries in each timestamp. Therefore, we don't need to make assumptions about the initial SOC of EVs and calculate the consumed energy based on trip distance. With the recorded SOC values, we can determine the consumed energy in trips and calculate the CD.

After applying all the data prepossessing procedures discussed above, we can derive the parameters of each trip, i.e., trip start time, end time, origin and destination (OD) locations, and consumed energy.

\vspace{-0.2cm}
\section{Travel behavior comparison}

The processed input data contains the records of 4,109 trips made by EVs during one year. This data can be used to extract the distribution of trip parameters, e.g., trip distance, trip start and end time, number of daily trips, etc., and presents an opportunity to compare the behavior of EVs and conventional ICEVs. Therefore, we repeated the same initial data processing on the dynamic trip data of ICEVs, and the result was the records of 18,936 trips made by ICEVs. We will use both of these data sets to compare the travel behavior between ICEVs and EVs.  

\begin{figure}
  \centering
  \subfloat[]{\includegraphics[width=\witwo in, height =\hetwo in]{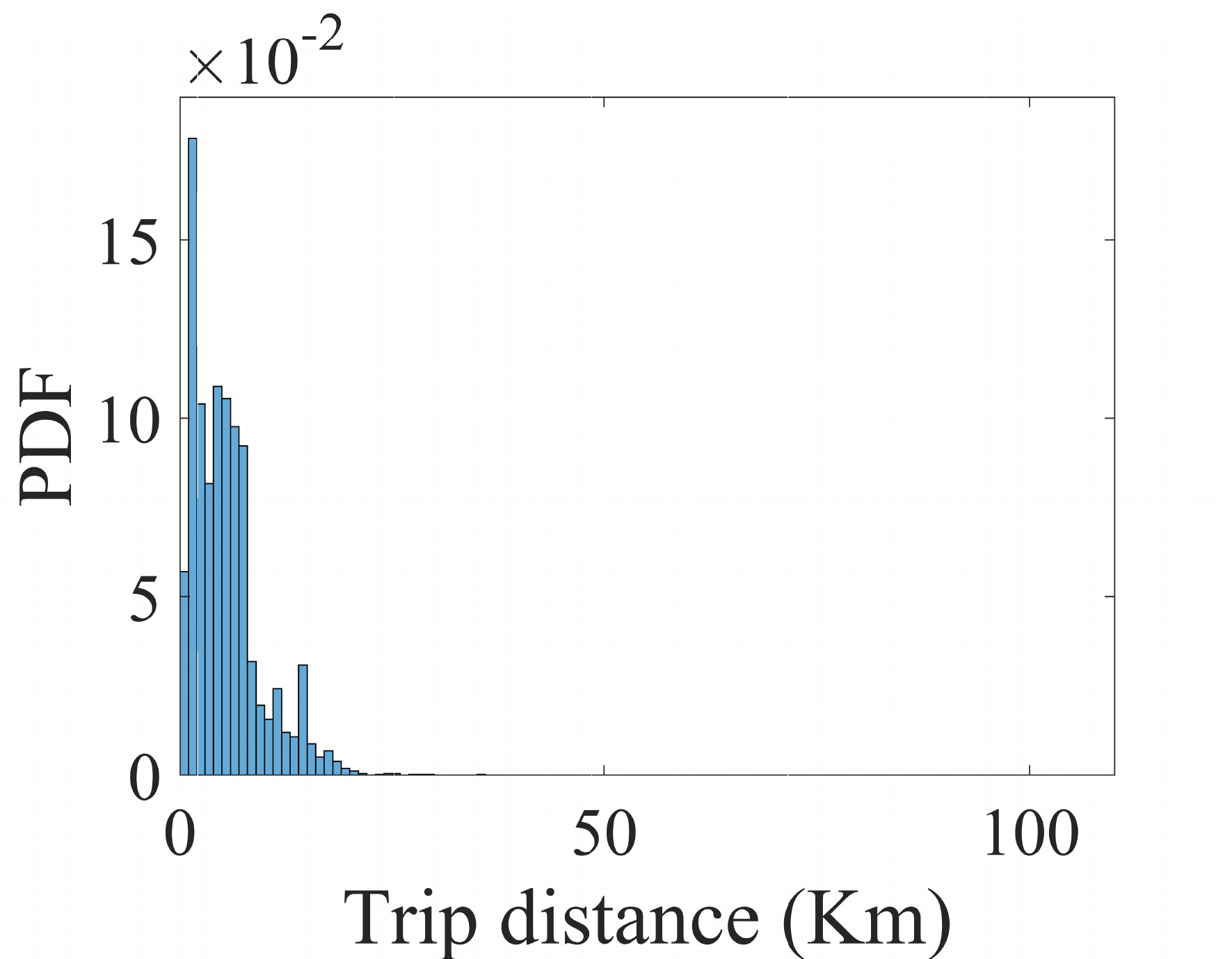}
  \label{fig:dist_EV}}
  \subfloat[]{\includegraphics[width=\witwo in, height =\hetwo in]{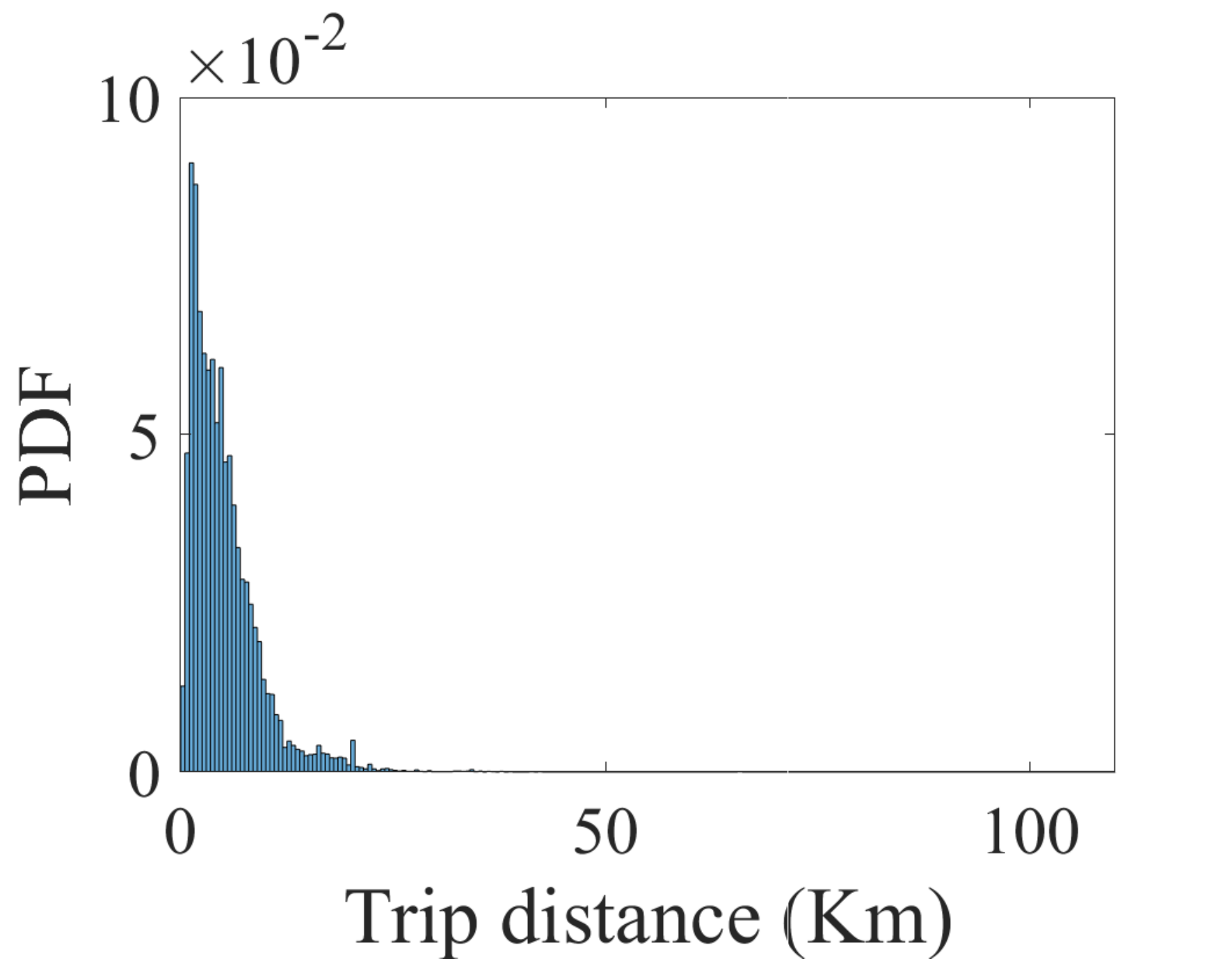}\label{fig:dist_ICE}}
      \captionsetup{justification=raggedright,singlelinecheck=false}
  \caption{Distribution of daily trips  (a) EVs (b) ICEVs} 
  \label{fig:dist_pdf}
\end{figure}

Fig. \ref{fig:dist_pdf} shows the distribution of daily trip distances for both EVs (Fig. \ref{fig:dist_EV}) and ICEVs (Fig.\ref{fig:dist_ICE}). The short distance trips (0$\sim$20 Km) are more prevalent in both types of vehicles, and the main difference is the maximum trip distance. ICEVs have a higher maximum trip distance (110 Km), where the maximum trip distance recorded for EVs is 35 Km. One can infer that ICEVs are preferred for long-distance trips. However, since the frequency of such trips is low, no general trends can be derived.  

\begin{figure}
  \centering
  \subfloat[]{\includegraphics[width=\witwo in, height =\hetwo in]{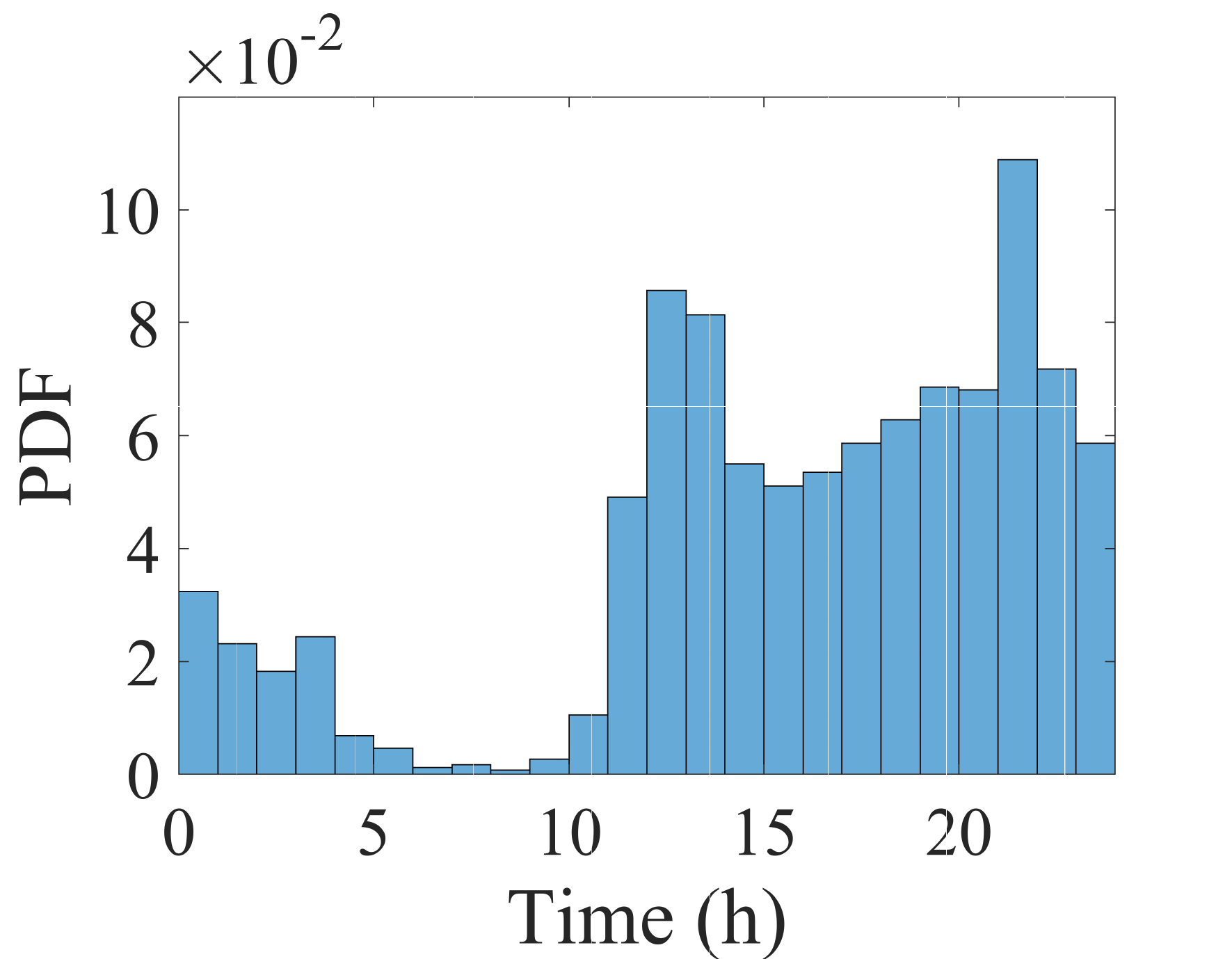}
  \label{fig:start_EV}}
  \subfloat[]{\includegraphics[width=\witwo in, height =\hetwo in]{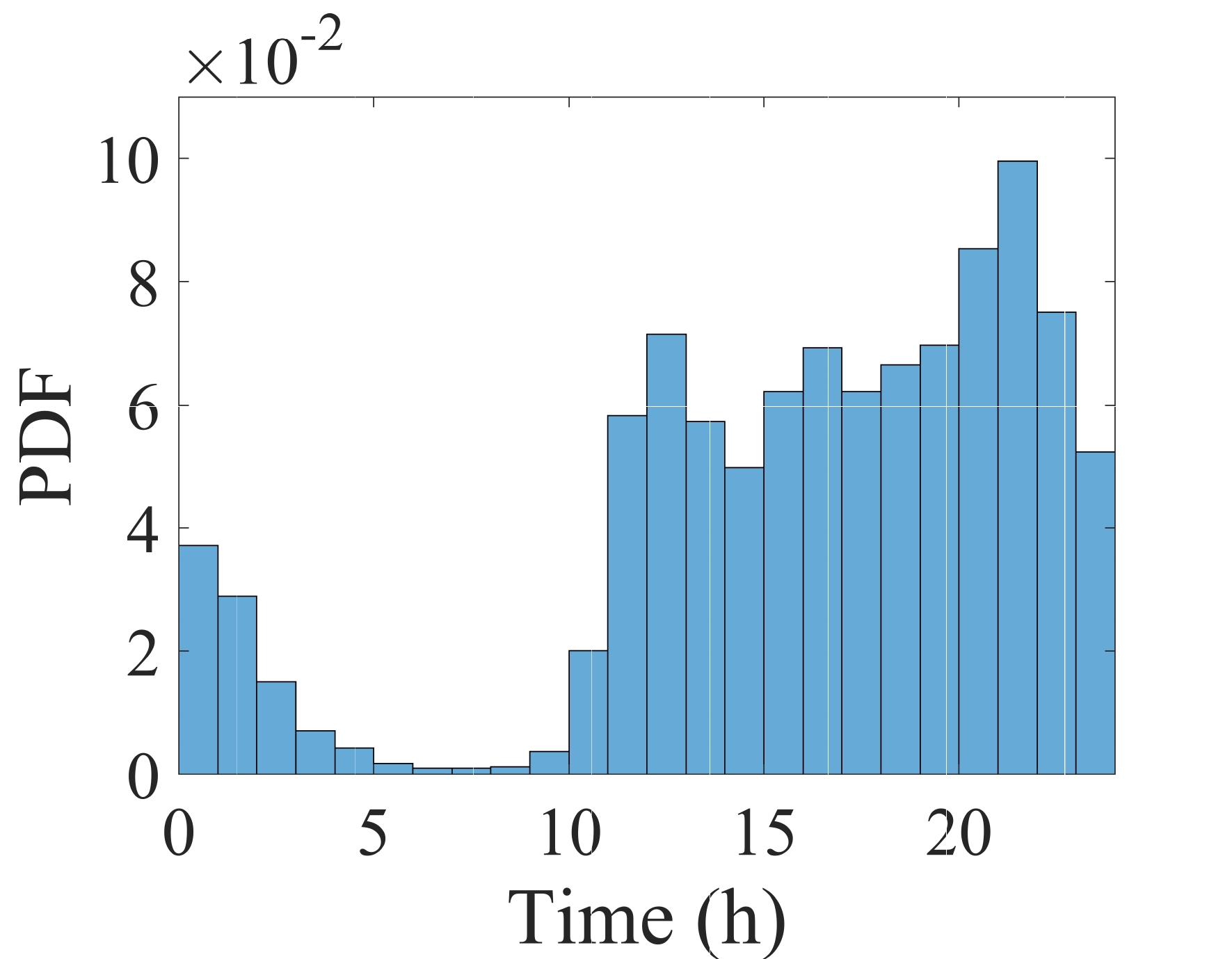}\label{fig:start_ICE}}
      \captionsetup{justification=raggedright,singlelinecheck=false}
  \caption{Distribution of daily trips' start time  (a) EVs (b) ICEVs} 
  \label{fig:start_time_pdf}
\end{figure}

\begin{figure}
  \centering
  \subfloat[]{\includegraphics[width=\witwo in, height =\hetwo in]{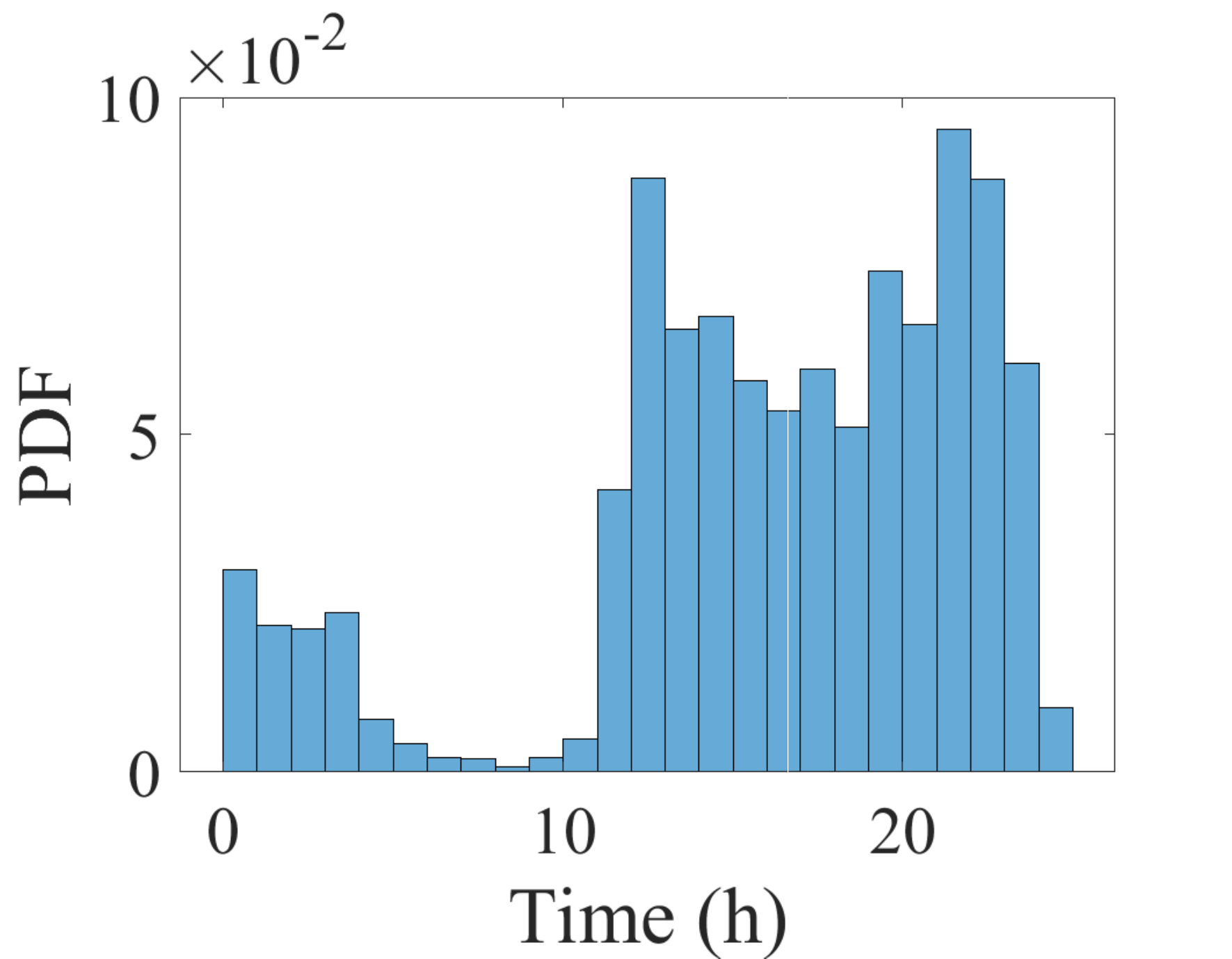}
  \label{fig:end_EV}}
  \subfloat[]{\includegraphics[width=\witwo in, height =\hetwo in]{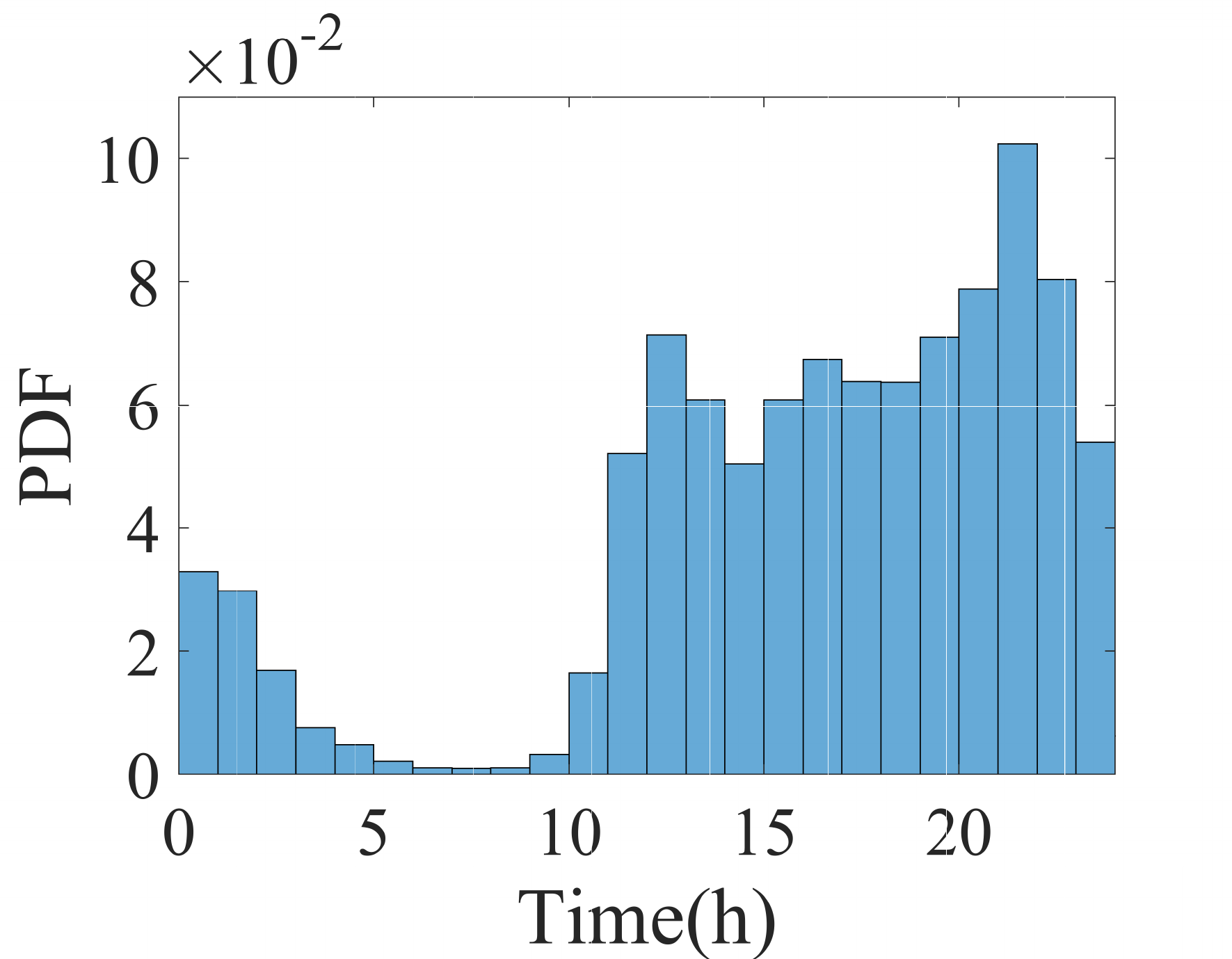}\label{fig:end_ICE}}
      \captionsetup{justification=raggedright,singlelinecheck=false}
  \caption{Distribution of daily trips' end time  (a) EVs (b) ICEVs} 
  \label{fig:end_time_pdf}
\end{figure}

The trips' start and end time distributions are shown in Fig. \ref{fig:start_time_pdf} and \ref{fig:end_time_pdf} for both types of vehicles. There are no major differences between the trip start time distributions for EVs and ICEVs (see Fig. \ref{fig:start_time_pdf}). Minor difference can be seen in the trip end time of EVs compared to ICEVs. EVs tend to finish their trips less during $t=23 \sim 1$ (see Fig. \ref{fig:end_EV}), whereas more trip end time has been recorded for ICEVs during that time (see Fig. \ref{fig:end_ICE}). Also, trip end time during $t=10 \sim 15$ has been more prevalent among EVs compared to ICEVs. This might be because of the charging needs of EVs after the trips or user preference of the drivers.  
% Another comparable parameter is the distribution of the number of daily trips made by the two types of vehicles. Fig. \ref{fig:trip_no} shows that EVs are more likely to make fewer trips per day compared to ICEVs. This could either due to the range anxiety that EV drivers or the current people buying EVs have lower travel demand (self-selection bias).

%the number of daily trips is similar for both EVs and ICEVs, and less probability for trip numbers 1$\sim$4 for ICEVs can be because the total trips recorded in the data for ICEVs (18,936 trips) is much more than the EV trips (4109 trips).

In summary, EVs and ICEVs have similar patterns on trip start/end time and distance per trip. But ICEVs owners may be more likely to have more daily trips and ICEVs are preferred for trips with long distances. Also, EVs tend to end their trip more during the afternoons and less close to the midnight.

% \begin{figure}
%   \centering
%   \includegraphics[width=0.45\linewidth]{Figures/trip_no.png}
%       \captionsetup{justification=raggedright,singlelinecheck=false}
%   \caption{Distribution of number of daily trips} 
%   \label{fig:trip_no}
% \end{figure}

\section{Charging Demand Modeling} \label{sec:demand_modeling}
In this section, we will derive the daily spatiotemporal CD behavior based on the input data. In order to detect the charging events, we will compare the SOC of the EVs at the end of each trip (${\mathrm{SOC}}^{\mathrm{arr}}_k$) and compare it to the start SOC of the next trip (${\mathrm{SOC}}^{\mathrm{dep}}_{k+1}$) with $k$ representing the trip index of the EV in the same day. If  ${\mathrm{SOC}}^{\mathrm{dep}}_{k+1} > {\mathrm{SOC}}^{\mathrm{arr}}_k$, trip $k$ would be a charging event starting at the end time of the same trip $k$. The CD can be calculated as the multiplication of the difference between SOCs for the two consecutive trips or the required SOC ($\mathrm{SOC}^\mathrm{req}$) and the battery capacity of each EV (${\mathrm{Cap}}_v$):
\begin{align}
    \mathrm{SOC}^\mathrm{req} = {\mathrm{SOC}}^{\mathrm{dep}}_{k+1} - {\mathrm{SOC}}^{\mathrm{arr}}_k\\
    {\mathrm{CD}}_k = \begin{cases} \mathrm{SOC}^\mathrm{req} \times {\mathrm{Cap}}_v & \mathrm{if} \; \mathrm{SOC}^\mathrm{req} > 0\\
    0 & \mathrm{Otherwise}
    \end{cases}\label{eq:demand}
\end{align}
The temporal behavior of CD also depends on the charging duration ($\Delta T$), i.e., the amount of time required to receive the demanded energy ($\mathrm{CD}_k$). This duration is directly proportional to the required CD and inversely proportional to the charging rate ($\alpha$) and charging efficiency ($\eta$), which depend on the battery characteristics and the installed charger. This relation is presented in (\ref{eq:ch_time}). We assumed that the demand ($\mathrm{CD}_k$) will be imposed on the system at the end time of trip $k$ lasting for a duration of $\Delta T_k$.
\begin{equation}
    \Delta T_k = \frac{{\mathrm{CD}}_k}{\alpha \cdot \eta} \label{eq:ch_time}
\end{equation}

\begin{figure}
  \centering
  \subfloat[]{\includegraphics[width=0.45\linewidth]{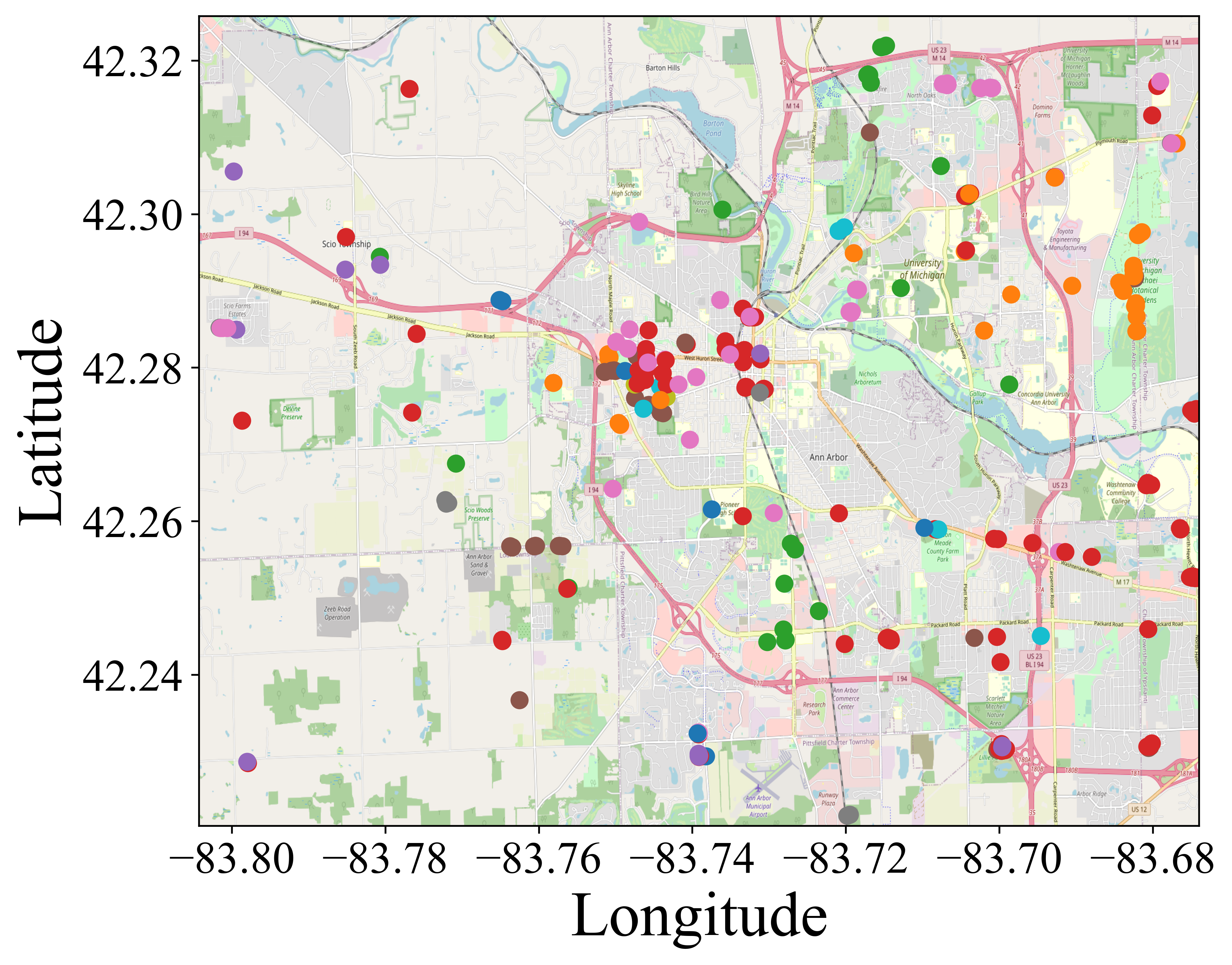}\label{fig:charging_loc}}
  \subfloat[]{\includegraphics[width=0.45\linewidth]{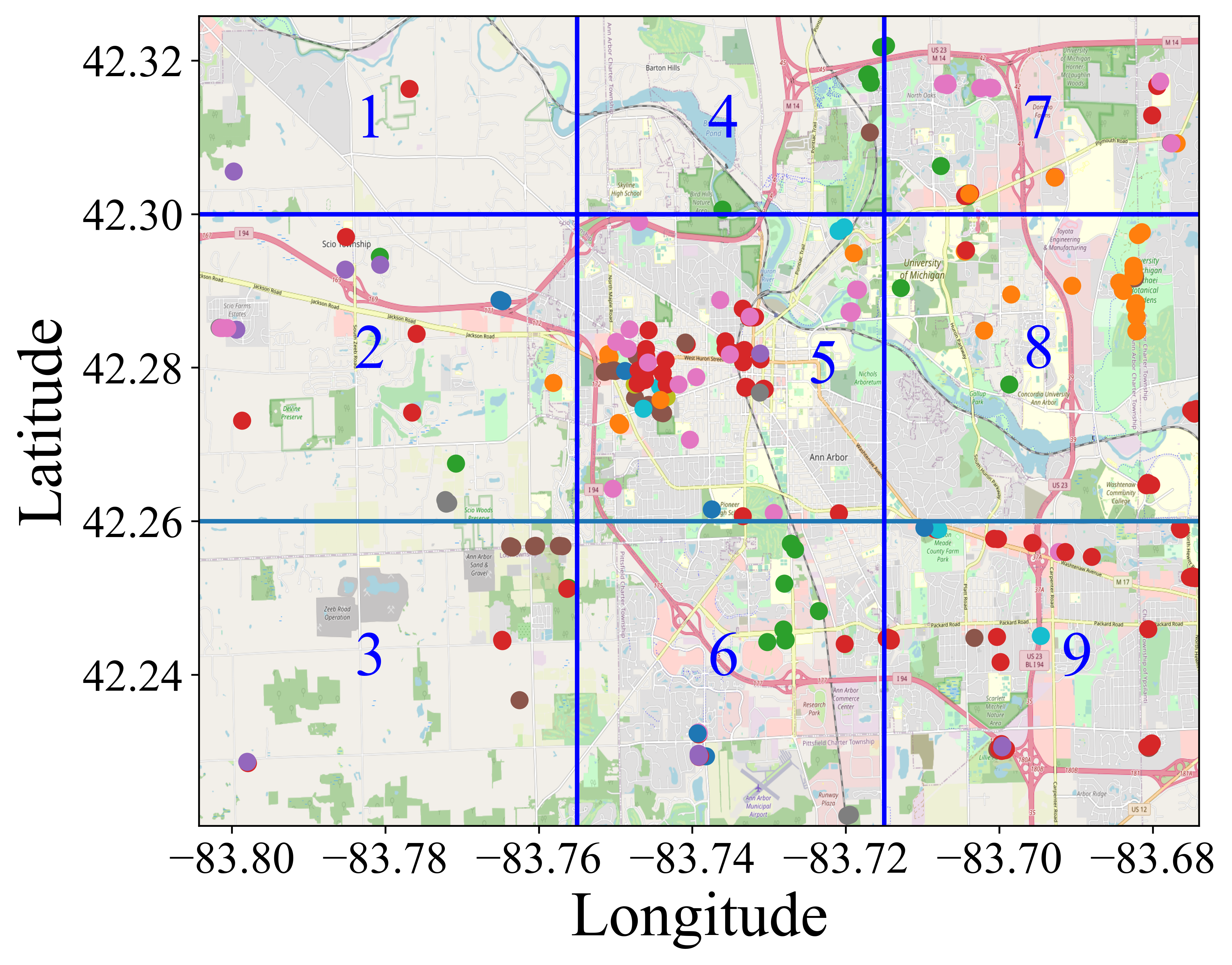}\label{fig:charging_zones}}
      \captionsetup{justification=raggedright,singlelinecheck=false}
  \caption{(a) Spatial distribution of charging events for different EVs and (b) considered charging zones of the city} 
\end{figure}

Additionally, the location of the charging event will be the destination of trip $k$, providing the spatial characteristics of the CD. Going through all the trips of EVs, we derived the charging location for each EV (see Fig. \ref{fig:charging_loc}) -- locations are color-coded for different EVs. We observe more charging instances located in the central part of the city compared to the other regions. The study region can be divided into finite number of zones $\{z_i\}_{1\leq i\leq n}$ covering different parts of the region, with $n$ being the total number of zones, to categorize the spatial distribution of charging events. The system operator (SO) can divide the region based on its requirements and judgments to any number of zones with various shapes; To ensure each zone has enough number of charging records, here we have considered to have $n$ = 9 charging zones as shown in Fig.\ref{fig:charging_zones}, which will be used to estimate the spatial location of the CD.

From the power system prospective, the SO can use the calculated CD and the expected charging location for operational purposes, e.g., day-ahead generation dispatch, real time energy management, etc. Additionally, the system planner can use the predicted spatial-temporal CD for the future expansion of charging infrastructure and estimate the required charging station capacities at different locations.

\section{Methodologies}
In this section, we will develop forecasting models to estimate the spatiotemporal pattern of daily CD. In addition, we will prepare the input data and define our input features and targets. 
\vspace{-0.25cm}
\subsection{Prepare Model Input Features}
At the start of each trip, the known parameters are a) trip start time, b) origin location, and c) start SOC (${\mathrm{SOC}}^{\mathrm{dep}}$). Our objective is to predict whether the EV will charge at the destination location; if yes, which zone and when this charging event will happen, and how much energy it will charge. 

Predicting the choice of charging at the end of a trip and the charging zone can be modeled as a classification problem. We will label each charging event trip of the input data based on the number of the zone of the event $L=i \; ({1\leq i \leq n})$, defined in Section \ref{sec:demand_modeling}, and define a dummy zone $L = {n+1}$ for the trips where no charging happens at the end of the trip. Table \ref{T:initial_data} shows how this labeling works on the example input data with $n$ = 9. Trips 2 and 4 are not considered as charging event trips because ${\mathrm{SOC}}^{\mathrm{req}} < 0$ and trip 4 is the last trip of the day with no information of the next trip's ${\mathrm{SOC}}^{\mathrm{dep}}$ in the next day (day 6). Therefore, the labels for these two trips are 10. Trips 1 and 3 are charging event trips with ${\mathrm{SOC}}^{\mathrm{req}} > 0$. The labels for these two trips are determined based on their destination locations, and they are in zones 9 and 8, respectively. 

The temporal behavior of this CD depends on the end time and charging duration and can be determined based on the procedure presented in Section \ref{sec:demand_modeling} using (\ref{eq:demand}) and (\ref{eq:ch_time}).

In summary, the input features of our problem are vehicle ID and current trip's start time, origin location, and ${\mathrm{SOC}}^{\mathrm{dep}}$ and we will use them to train the forecasting models. 

\begin{table}
\centering
\renewcommand{\arraystretch}{1}
\footnotesize
\captionsetup{labelsep=space,font={footnotesize,sc}}
\caption{Processed data example for the forecasting models}\label{T:initial_data}
\centering
\resizebox{0.9\columnwidth}{!}{
\centering
\begin{tabular}{|c|c|c|c|c|c|c|c|c|}
\hline\hline
\multirow{2}{*}{\tabincell{c}{Vehicle \\ ID}} & \multirow{2}{*}{\tabincell{c}{Trip \\ No.}} & \multirow{2}{*}{\tabincell{c}{$T^{\mathrm{str}}$ \\ (h)}} & \multirow{2}{*}{\tabincell{c}{$T^{\mathrm{end}}$ \\ (h)}} & \multicolumn{2}{c|}{\tabincell{c}{Origin location}} & \multirow{2}{*}{\tabincell{c}{Label}} & \multirow{2}{*}{\tabincell{c}{${\mathrm{SOC}}^{\mathrm{dep}}$\\ $\%$}} & \multirow{2}{*}{\tabincell{c}{${\mathrm{SOC}}^{\mathrm{req}}$\\ $\%$}}\\ \cline{5-6}
                     &                    &                    &                    &      Latitude          & Longitude         &           & &                 \\ \hline
 371                &  1                  &  13.44     &  13.75 &   42.277  & -83.75 &   9   &   40.92    &  41.96      \\ 
 371                &   2         &    18.47     & 18.56    & 42.253    & -83.674   & 10 &  45.69 & 0         \\ 
371                &  3           &  18.96   &  19.06     &   42.256    &  -83.696   &   8  &  33.13 & 46.99     \\ 
371                & 4              &  15.45   &  15.57   &    42.302   &  -83.704   &  10    &  64.67 & 0 \\ 
\hline\hline
                            
\end{tabular}}
\end{table}
\vspace{-0.2cm}
\subsection{Forecasting Models}

The forecasting targets in our modeling are the label of the trip, trip end time, and ${\mathrm{SOC}}^{\mathrm{Req}}$. Table \ref{T:inputs_targets} shows the input features and the targets based on Table \ref{T:initial_data} as an example. Note that we calculate the CD only for trips with charging events, so we need to add charging labels as an input feature for forecasting the CD.

% Before feeding the input data to the forecasting models, we extract the trips made in days with charging event occurrence (1,062 trips) and split them into training sets and test sets -- 25 $\%$ test set and 75$\%$ training sets. The next step is to develop forecasting models and predict each of the defined targets. 

\begin{table}
\centering
\renewcommand{\arraystretch}{0.9}
\footnotesize
\captionsetup{labelsep=space,font={footnotesize,sc}}
\caption{Input features and targets of the example data}\label{T:inputs_targets}
\centering
\resizebox{0.9\columnwidth}{!}{
\begin{tabular}{|c|c|c|c|c|c|c|c|c|}
\hline \hline
\multicolumn{5}{|c|}{\multirow{1}{*}{\tabincell{c}{Input features}}} & \multicolumn{3}{c|}{\multirow{1}{*}{\tabincell{c}{Targets}}} \\ 
\hline
{\multirow{2}{*}{\tabincell{c}{Vehicle  \\ ID}}} &
{\multirow{2}{*}{\tabincell{c}{$T^{\mathrm{str}}$  \\ (h)}}} & {\multirow{2}{*}{\tabincell{c}{${\mathrm{SOC}}^{\mathrm{dep}}$  \\ $\%$}}} & \multicolumn{2}{c|}{\multirow{1}{*}{{{Origin location}}}}  & 
{\multirow{2}{*}{\tabincell{c}{Label}}}& 
{\multirow{2}{*}{\tabincell{c}{$T^{\mathrm{end}}$  \\ (h)}}} & {\multirow{2}{*}{\tabincell{c}{${\mathrm{SOC}}^{\mathrm{req}}$  \\ $\%$}}}  
\\  \cline{4-5}
 & & & Latitude & Longitude &  & & \\ \hline
371   & 13.44  & 40.92   & 42.277   & -83.75 & 9   & 13.75  & 41.96    \\

371   & 18.47 & 45.69   & 42.253  & -83.674  & 10  &  18.56   & 0       \\

371  & 18.96  &  33.13 & 42.256 & -83.696  & 8  & 19.06   & 46.99        \\

371  & 15.45  & 64.67  & 42.302  &  -83.704   & 10   &  15.57  &  0  \\
\hline\hline
                            
\end{tabular}
}
\end{table}

Five forecasting methods, namely K-nearest neighbor (KNN), decision tree (DT), random forest (RF), artificial neural networks (ANNs), and deep artificial neural networks (DANNs) were considered to predict each target. The considered methods have been widely used in machine learning applications and the detailed comparison on these methods are presented in \cite{jadhav2016comparative}. We will train the models and determine their hyper parameters separately for each target with the defined input features. These parameters include the number of neighbors for the KNN method, tree depth for DT and RF methods, number of neurons for ANN, and number of neurons on each hidden layer for the DANN method. We investigated the performance of the models for a range of parameters by observing the accuracy of the classification for trip label prediction and the forecasting error for the trip end time and required SOC. The selected parameters and performance evaluation of the forecasting models are discussed in Section \ref{sec:results}. 

% We evaluated the forecasting performance of the methods on test sets and training sets and selected the best parameters for each target based on the best performance on the test data set (see Table \ref{T:model_performance}). Table \ref{T:model_parameters} summarizes the selected parameters for different forecasting methods. For the DANN method, the number of hidden layers more than the mentioned layers in Table \ref{T:model_parameters} resulted in less accuracy. Therefore, we limited the number of layers to 2 for CS zone (label) and trip end time prediction, and 3 layers for ${\mathrm{SOC}^{\mathrm{req}}}$ prediction. 

% evaluated the forecasting performance of the methods on test sets and training sets and selected the best parameters for each target based on the best performance on the test data set (see Table \ref{T:model_performance}).
\vspace{-0.3cm}
\section{Results}\label{sec:results}
\vspace{-0.2cm}
\subsection{Parameter Settings}

Before feeding the input data to the forecasting models, we extract the trips made in days with charging event occurrence (1,062 trips) and split them into training sets and test sets -- 25 $\%$ test set and 75$\%$ training sets. Scickit-learn library was used to train the KNN, DT, and RF models, and TensorFlow library were used to train the ANN based models. All the implementations were done on the Python software. 

We trained the forecasting methods for a range of modeling parameters using the training sets and observed the performance of the methods on test set to select the best modeling parameters. Table \ref{T:model_parameters} summarizes the selected parameters for different forecasting methods. For the DANN method, the number of hidden layers more than the mentioned layers in Table \ref{T:model_parameters} resulted in less accuracy. Therefore, we limited the number of layers to 2 for CS zone (label) and trip end time prediction, and 3 layers for ${\mathrm{SOC}^{\mathrm{req}}}$ prediction.

\begin{table}
\centering
\renewcommand{\arraystretch}{1.1}
\footnotesize
\captionsetup{labelsep=space,font={footnotesize,sc}}
\caption{Forecasting model parameters}\label{T:model_parameters}
\centering
\resizebox{0.9\columnwidth}{!}{\begin{tabular}{|c|c|c|c|c|c|c|}
\hline\hline
\multirow{2}{*}{Target} & \multirow{2}{*}{\tabincell{c}{KNN \\ neighbors}} & \multirow{2}{*}{\tabincell{c}{DT \\ tree depth}} & \multirow{2}{*}{\tabincell{c}{RF\\ tree depth}} & \multirow{2}{*}{\tabincell{c}{ANN\\ neurons}}&  \multicolumn{2}{c|}{\multirow{1}{*}{\tabincell{c}{DANN}}} \\ \cline{6-7}
& & & & & Layers & Neurons \\
\hline
\multirow{1}{*}{\tabincell{c}{Label}}  &  \multirow{1}{*}{11}   &  \multirow{1}{*}{8}  &  \multirow{1}{*}{6} & \multirow{1}{*}{400} & 2 & 500,100 \\ \hline
\multirow{1}{*}{\tabincell{c}{$T^{\mathrm{end}}$}} &  \multirow{1}{*}{4}   &  \multirow{1}{*}{8}  &  \multirow{1}{*}{18} & \multirow{1}{*}{500} & 2 & 400,100\\  \hline
\multirow{1}{*}{${\mathrm{SOC}}^{\mathrm{req}}$}  &  \multirow{1}{*}{10}   &  \multirow{1}{*}{4}  &  \multirow{1}{*}{5} & \multirow{1}{*}{400} &3 & 900,500,100\\ 
\hline \hline
\end{tabular}}
\end{table}

According to \cite{prehofer2020big}, the battery capacities of EVs studied in the VED data set is in the range of 20$\sim$25 KWh. Therefore, we consider the battery characteristics of Fiat500E, which is reported as one of the popular EVs in Alternative Fuels  Data  Center \cite{AFDC2020} and has a battery capacity of 24 KWh. We assume a charging rate of  $\alpha = 6.6$ KW (i.e., level 2 charging) and the charging efficiency to be $\eta = 0.9$ \cite{AFDC2020} for CD demand calculations mentioned in (\ref{eq:ch_time}). 
\vspace{-0.3cm}
\subsection{Charging Demand Forecasting}

After determining the parameters for each forecasting model, we compared their performance for predicting each target on the test data set (see Table \ref{T:model_performance}). The RF method provided the highest accuracy for charging zone label prediction and the least RMSE value for the trip end time and the DT method had the least RMSE value for predicting the required SOC. Even though ANN-based models have provided superior performance in other machine learning studies \cite{jahangir2019charging,jahangir2019novel}, simpler models (KNN, DT, and RF) provided better results for our dataset. The low accuracy of label prediction stems from the small number of charging event trips (446 trips) compared to all of the trips made in days with charging events (1,062 trips) and a large number of classes/zones (10 classes). 

\begin{table}
\centering
\renewcommand{\arraystretch}{1.1}
\footnotesize
\captionsetup{labelsep=space,font={footnotesize,sc}}
\caption{Performance of forecasting models on the test set: classifying accuracy for label prediction and RMSE value for $T^{\mathrm{end}}$ and ${\mathrm{SOC}}^{\mathrm{req}}$}\label{T:model_performance}
\centering
\resizebox{0.8\columnwidth}{!}{\begin{tabular}{|c|c|c|c|c|c|}
\hline\hline
{Target} &\tabincell{c}{KNN} &\tabincell{c}{DT} & \tabincell{c}{RF} & ANN & DANN \\ \hline
{\tabincell{c}{Label ($\%$)}}  &  63.91   &  63.53  &  73.97 & 65.73 & 65.78   \\ \hline
{\tabincell{c}{$T^{\mathrm{end}}$ (h)}} &  0.39   &  0.17  &  0.15 & 0.19& 0.19 \\ \hline
{${\mathrm{SOC}}^{\mathrm{req}}$ ($\%$)}  &  17.05   &  15.55  &  15.79 & 21.73 & 17.57 \\ \hline \hline
\end{tabular}}

\end{table}

\begin{figure}
  \centering
  \subfloat[]{\includegraphics[width=\witwo in, height =\hetwo in]{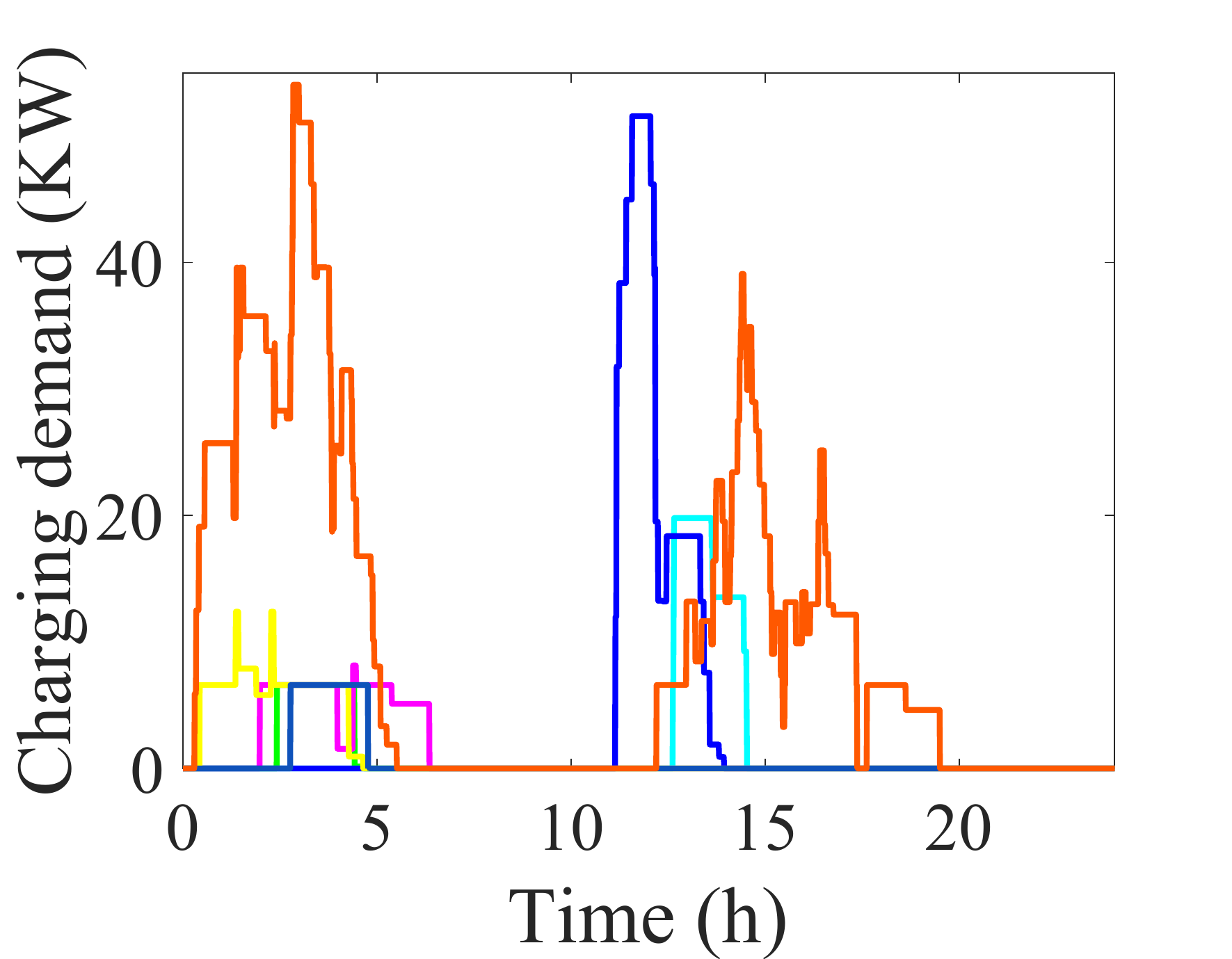}\label{fig:CD_1}}
  \subfloat[]{\includegraphics[width=\witwo in, height =\hetwo in]{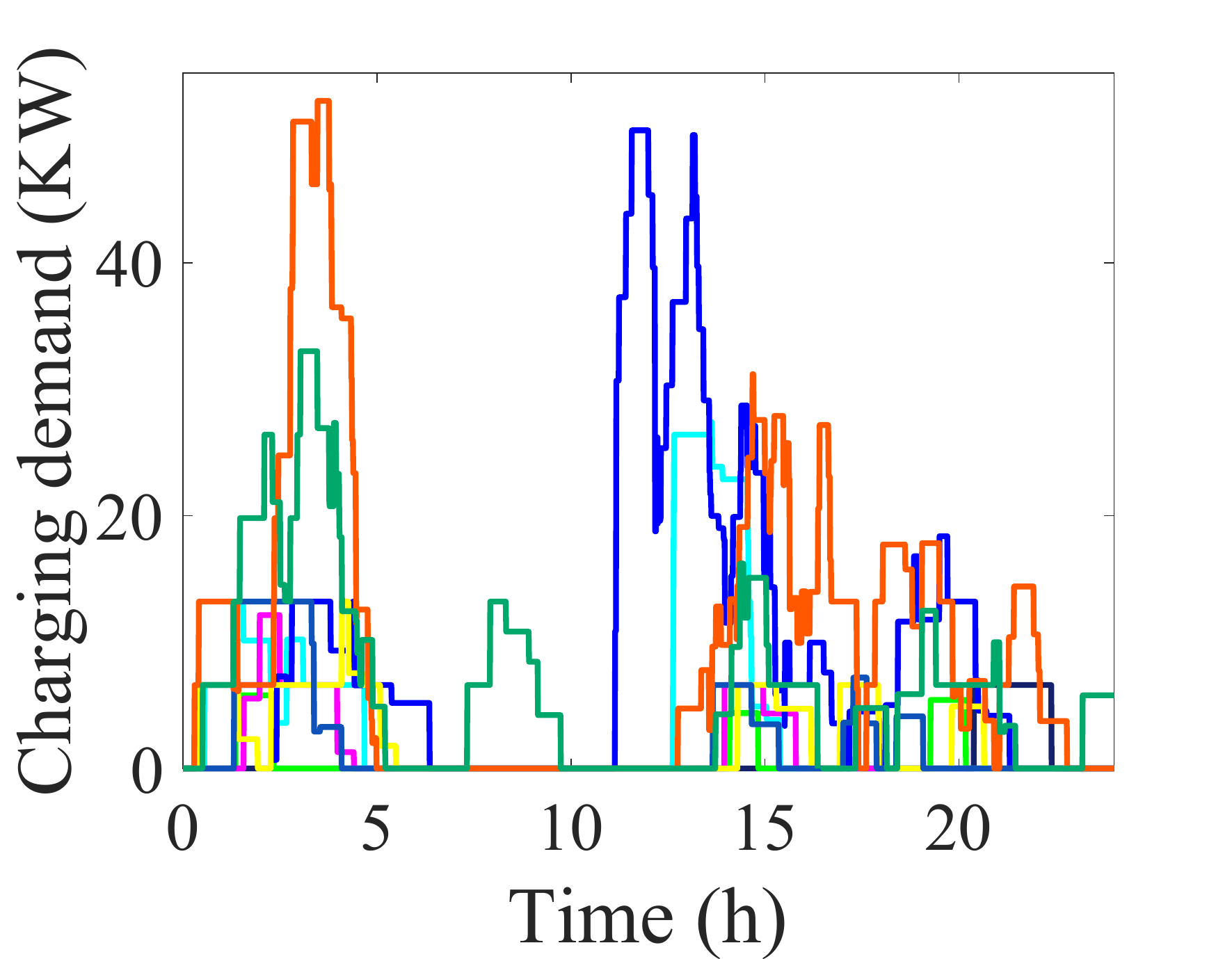}\label{fig:CD_2}}
  \vspace{-0.4cm}
  \subfloat[]{\includegraphics[width=\witwo in, height =\hetwo in]{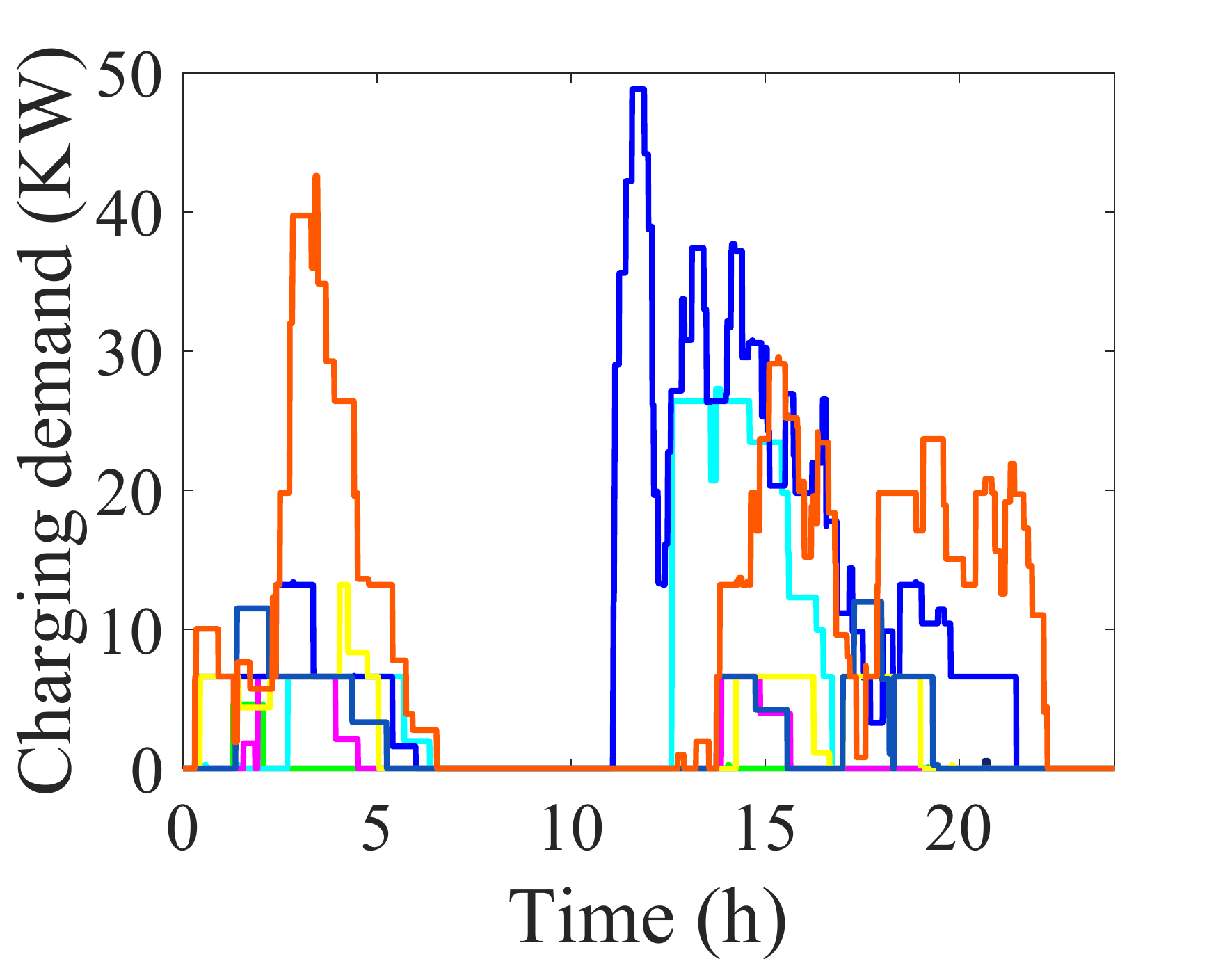}\label{fig:CD_test}}
  \subfloat[]{\includegraphics[width=\witwo in, height =\hetwo in]{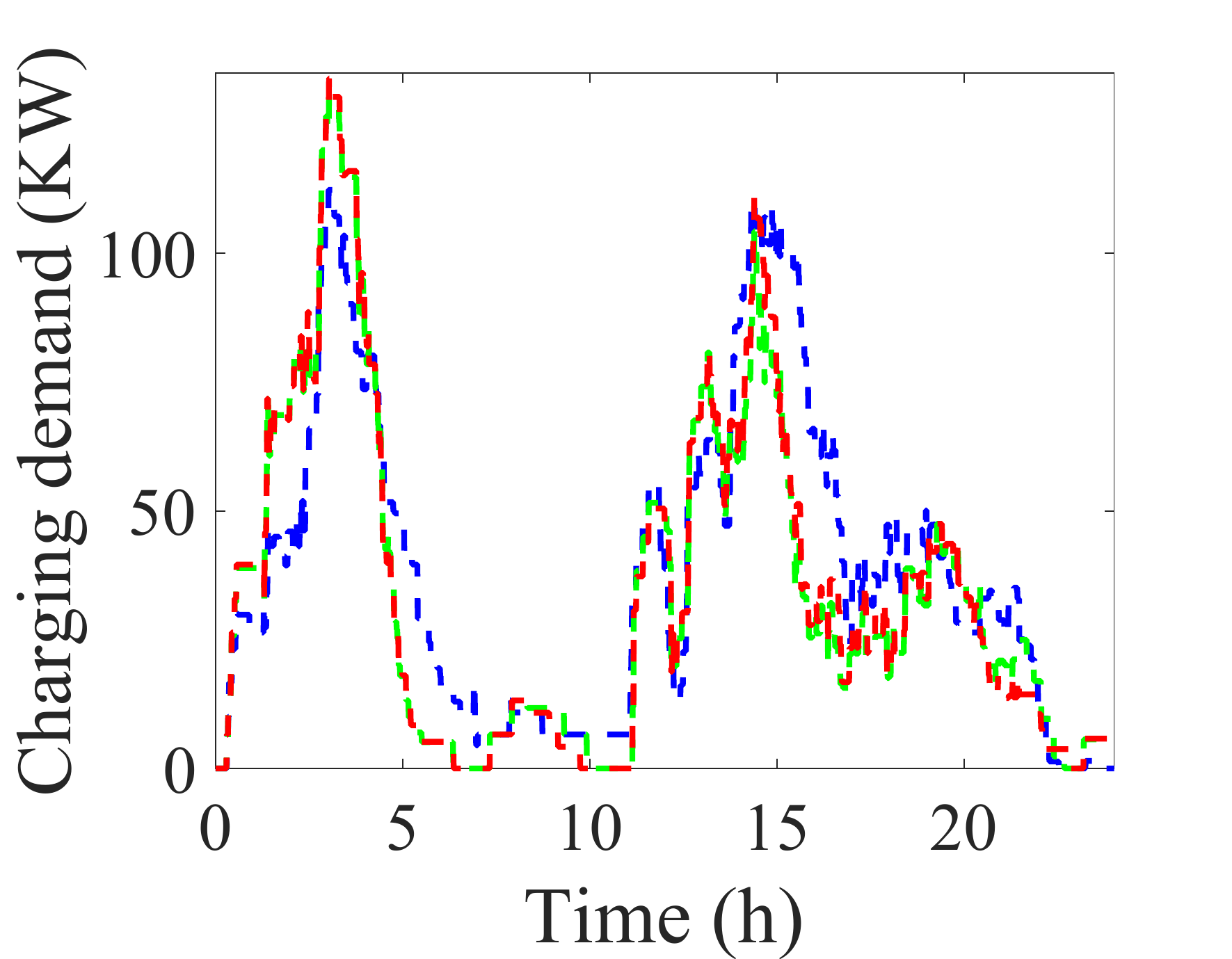}\label{fig:CD_total_estimated}}
    \captionsetup{justification=raggedright,singlelinecheck=false}
  \caption{Zonal CD for (a) case 1 (b) case 2 (c) test data, and (d) total CD; \textcolor{black}{\textbf{---}} zone 1, \textcolor{um}{\textbf{---}} zone 2, \textcolor{green}{\textbf{---}} zone 3, \textcolor{c}{\textbf{---}} zone 4, \textcolor{blue}{\textbf{---}} zone 5, \textcolor{m}{\textbf{---}} zone 6, \textcolor{yellow}{\textbf{---}} zone 7, \textcolor{orange}{\textbf{---}} zone 8, \textcolor{Abi}{\textbf{---}} zone 9; \textcolor{blue}{\textbf{-.-}} Test data, \textcolor{green}{\textbf{-.-}} Estimated case 1, \textcolor{red}{\textbf{-.-}} Estimated case 2. }\label{fig:zonal_estimated_demands}
  
\end{figure}

We selected the best forecasting models for each target and predicted the spatiotemporal pattern of daily CD (see Figure \ref{fig:zonal_estimated_demands}). We considered two cases for forecasting the $\mathrm{SOC}^{\mathrm{req}}$ 1) using the predicted labels as the input feature of the trained DT model and 2) using the actual labels of the trips. Charging start times were based on the predicted trip end times, and the resulting daily CDs were calculated using the procedure delineated in Section \ref{sec:demand_modeling}.

Comparing the estimated demands from case 1 and the test data (see Fig. \ref{fig:CD_1} and \ref{fig:CD_test}) shows that the developed classifier failed to detect the charging events in zone 1. This discrepancy is because of the less recorded charging events in this zone (see Fig. \ref{fig:charging_zones}). On the contrary, most of the charging events are predicted to occur at zone 5, which has more charging records in the input data, causing a significant spike in CD of this zone at $t=11\sim12$.  Moreover, the estimated CD for case 1 follows the temporal pattern of the CD during $t=0\sim6$ but performs poorly in the afternoon (see Fig. \ref{fig:CD_1}) because more charging events are predicted to be at zone 5 and most of the charging events in this zone happen in $t=11\sim16$.

Passing the actual zones from the test data for forecasting the ${\mathrm{SOC}}^{\mathrm{req}}$ (case 2) improved the temporal pattern in the afternoon and the nighttime (see Fig. \ref{fig:CD_2}). The accurate charging zones have reduced the charging events in zone 5 and we no longer observe the high peak CD during $t=11\sim12$ (compare Fig.\ref{fig:CD_1} and \ref{fig:CD_2}). This increase in accuracy points out the importance of charging zone prediction in determining CD's temporal and spatial behavior.   

For comparing the temporal pattern of CD, we have derived the total daily CD (see Fig. \ref{fig:CD_total_estimated}). The estimated CD after midnight until the morning  ($t=0\sim6$) follows the same pattern in both cases and resembles the actual CD derived from test data. However, in the afternoon, the prediction inaccuracy in charging zones resulted in inaccurate temporal estimates for case 1, especially after $t=17$. 

The SO and system planners can use the estimated spatio-temporal CD in different applications: 1) The predicted zonal CD in Fig. \ref{fig:CD_2} can be used for day ahead generation dispatch and other operational planning purposes. For example, it is estimated that the CD at zone 8 would be high during the morning. Therefore, the SO would consider allotting sufficient energy generation in the day ahead planning of that zone. 2) The system planner can use the expected zonal CD for inferring user preferences and expanding the CS infrastructures. For example, the daily CD pattern shows higher CD at zones 5 and 8 compared to the other zones (see Fig. \ref{fig:CD_2}), which can be good candidates for CS expansion. 
% \begin{figure}
%   \centering
%   \includegraphics[width=0.4\linewidth]{Figures/total_demand_estimated.png}
%       \captionsetup{justification=raggedright,singlelinecheck=false}
%   \caption{Total CD estimated and derived from the test data} 
  
% \end{figure}

% VED contains high-resolution GPS trajectories of both EVs and conventional vehicles, which allows us to compare the travel behavior of different vehicle types. 
\vspace{-0.3cm}
\section{Summary and Future Extensions}
Estimating EVs' daily CD's spatial and temporal pattern is a vital factor in modeling uncertainties in power systems. CD pattern is closely related to the travel behavior of EVs. In this study, we use the VED, a high-resolution GPS data set of both EVs and conventional vehicles, to investigate the potential differences in the travel behavior of EVs and conventional vehicles and analyzed the real spatiotemporal pattern of CD.  Our analyses showed that the travel behavior of EVs is similar to ICEVs in terms of trip start and end time, and trip distance. But EVs tend to have fewer daily trips compared to ICEVs. 

After processing the GPS trajectories of EVs, we developed forecasting models to predict the spatiotemporal pattern of CD based on the known information at the starting point of the trips. The spatial pattern of CD was modeled as a classification problem, where the charging event locations were grouped in different zones. The CD and its temporal pattern were estimated using a decision tree and random forest regressors with high accuracy. 

% However, the classification model showed less accurate results in predicting the zones of charging events due to the limited volume of the input data.

The small scale of the input data limited the performance of the forecasting model, and we will examine the models on larger data sets as more data sets are becoming available for EVs. Future extension should also investigate the potential reasons of different predicting accuracy to better understand the applicability of machine learning algorithms. Another extension of this work is to develop a forecasting model to predict the daily CD based on the historical travel data rather than based on information at the starting point of each trip.
%, which was lacking in the VED dataset where trip records are missing.
% \newpage
\vspace{-0.3cm}
\setstretch{0.9}
\bibliographystyle{./bibliography/IEEEtran}	
\bibliography{ref.bib}

\end{document}